\documentclass[11pt]{article}
\usepackage{graphicx}
\usepackage[margin=1.25in]{geometry}
\usepackage[usenames,dvipsnames]{color}
\usepackage{url}
\usepackage[colorlinks = true,
            linkcolor = blue,
            urlcolor  = blue,
            citecolor = blue,
            anchorcolor = blue]{hyperref}

\newcommand\pubnumber{FERMILAB-PUB-22-241-SQMS-TD}
\newcommand\pubdate{\today}

\def\Title#1{\begin{center} {\LARGE #1 } \end{center}}
\def\Author#1{\begin{center}{ \sc #1} \end{center}}
\def\Address#1{\begin{center}{ \it #1} \end{center}}

\newcommand\pubblock{\rightline{\begin{tabular}{l} \pubnumber\\
         \pubdate \end{tabular}}}
\newenvironment{Abstract}{\begin{quotation} \begin{center}
                       ABSTRACT
     \end{center}\bigskip  }{\end{quotation}}

\def\beq{\begin{equation}}
\def\eeq#1{\label{#1}\end{equation}}
\def\eeqn{\end{equation}}

\newenvironment{Eqnarray}%
   {\arraycolsep 0.14em\begin{eqnarray}}{\end{eqnarray}}
\def\beqa{\begin{Eqnarray}}
\def\eeqa#1{\label{#1}\end{Eqnarray}}
\def\eeqan{\end{Eqnarray}}

\let\bar=\overbar

\def\lsim{\mathrel{\raise.3ex\hbox{$<$\kern-.75em\lower1ex\hbox{$\sim$}}}}
\def\gsim{\mathrel{\raise.3ex\hbox{$>$\kern-.75em\lower1ex\hbox{$\sim$}}}}

\def\del{\partial}
\def\Dslash{\not{\hbox{\kern-4pt $D$}}}
\def\dslash{\not{\hbox{\kern-2pt $\del$}}}
\def\pslash{\not{\hbox{\kern-2pt $p$}}}
\def\ETmiss{\not{\hbox{\kern-4pt $E$}}_T}

\def\Dlr{\mathrel{\raise1.5ex\hbox{$\leftrightarrow$\kern-1em\lower1.5ex\hbox{$D$}}}}

\def\MSB{{\bar{M \kern -2pt S}}}
\def\msb{{\bar{\scriptsize M \kern -1pt S}}}

\def\drb{{\bar{\scriptsize D \kern -1pt R}}}

\newcommand\snowmass{\begin{center}\rule[-0.2in]{\hsize}{0.01in}\\\rule{\hsize}{0.01in}\\
\vskip 0.1in Submitted to the  Proceedings of the US Community Study\\ 
on the Future of Particle Physics (Snowmass 2021)\\ 
\rule{\hsize}{0.01in}\\\rule[+0.2in]{\hsize}{0.01in} \end{center}}

\begin{document}

\pubblock

\Title{\textbf{Key directions for research and development of superconducting radio frequency cavities}}
\medskip

\begin{Abstract}
\noindent
Radio frequency superconductivity is a cornerstone technology for many future HEP particle accelerators and experiments from colliders to proton drivers for neutrino facilities to searches for dark matter. While the performance of superconducting radio frequency (SRF) cavities has improved significantly over the last decades, and the SRF technology has enabled new applications, the proposed HEP facilities and experiments pose new challenges. To address these challenges, the field continues to generate new ideas and there seems to be a vast room for improvements. In this paper we discuss the key research directions that are aligned with and address the future HEP needs.
\end{Abstract}

\snowmass

\medskip 

\Author{\textbf{S.~Belomestnykh}$^{1,16,*}$ and \textbf{S.~Posen}$^{1,*}$}

\Author{D.~Bafia$^1$, S.~Balachandran$^{12}$, M.~Bertucci$^{10}$, A.~Burrill$^{15}$, A.~Cano$^1$, M.~Checchin$^1$, G.~Ciovati$^{19}$, L.D.~Cooley$^{12}$, G.~Dalla Lana Semione$^9$, J.~Delayen$^{14,19}$, G.~Eremeev$^1$, F.~Furuta$^1$, F.~Gerigk$^5$, B.~Giaccone$^1$, D.~Gonnella$^{15}$, A.~Grassellino$^1$, A.~Gurevich$^{14}$, W.~Hillert$^9$, M.~Iavarone$^{18}$, J.~Knobloch$^{7,21}$, T.~Kubo$^{11,17}$, W.-K.~Kwok$^2$, R.~Laxdal$^{20}$, P.J.~Lee$^{12}$, M.~Liepe$^3$, M.~Martinello$^1$, O.S.~Melnychuk$^1$, A.~Nassiri$^2$, A.~Netepenko$^1$, H.~Padamsee$^{3,1}$, C.~Pagani$^{10}$, R.~Paparella$^{10}$, U.~Pudasaini$^{19}$, C.E.~Reece$^{19}$, D.~Reschke$^4$, A.~Romanenko$^1$, M.~Ross$^{15}$, K.~Saito$^6$, J.~Sauls$^{13}$, D.N.~Seidman$^{13}$, N.~Solyak$^1$, Z.~Sung$^1$, K.~Umemori$^{11}$, A.-M.~Valente-Feliciano$^{19}$, W.~Venturini~Delsolaro$^5$, N.~Walker$^4$, H.~Weise$^4$, U.~Welp$^2$, M.~Wenskat$^9$, G.~Wu$^1$, X.X.~Xi$^{18}$, V.~Yakovlev$^1$, A.~Yamamoto$^{11,5}$, and J.~Zasadzinski$^8$}


\Address{
$^1$Fermi National Accelerator Laboratory, Batavia, IL 60510, USA \protect\\
$^2$Argonne National Laboratory, Argonne, IL, 60439, USA \protect\\
$^3$Cornell University, Ithaca, NY 14853, USA \protect\\
$^4$Deutsches Elektronen-Synchrotron, Notkestrasse 85, 22607 Hamburg, Germany \protect\\
$^5$European Organization for Nuclear Research (CERN), Geneva, Switzerland \protect\\
$^6$Facility for Rare Isotope Beams, Michigan State University, East Lansing, MI, 48824, USA \protect\\
$^7$Helmholtz-Zentrum Berlin, Albert-Einstein-Str. 15, 12489 Berlin, Germany \protect\\
$^8$Illinois Institute of Technology, Chicago, Illinois 60616, USA \protect\\
$^9$Institute of Experimental Physics, University of Hamburg, Luruper Chaussee 149, 22761 Hamburg, Germany \protect\\
$^{10}$Istituto Nazionale di Fisica Nucleare (INFN) LASA, Segrate, Italy \protect\\
$^{11}$High Energy Accelerator Research Organization (KEK), Tsukuba, Ibaraki 305-0801, Japan \protect\\
$^{12}$National High Magnetic Field Laboratory, Tallahassee, FL 32310, USA\protect\\
$^{13}$Northwestern University, Evanston, IL 60208, USA \protect\\
$^{14}$Old Dominion University, Norfolk, VA 23529 USA \protect\\
$^{15}$SLAC National Accelerator Laboratory, Menlo Park, CA 94025, USA \protect\\
$^{16}$Stony Brook University, Stony Brook, NY 11794, USA \protect\\
$^{17}$The Graduate University for Advanced Studies (Sokendai), Hayama, Kanagawa 240-0193, Japan \protect\\
$^{18}$Temple University, Philadelphia, PA 19122 \protect\\
$^{19}$Thomas Jefferson National Accelerator Facility, Newport News, VA 23606, USA \protect\\
$^{20}$TRIUMF, Vancouver, British Columbia, V6T2A3, Canada \protect\\
$^{21}$Universität Siegen, Walter-Flex-Str. 3, 57068 Siegen, Germany
}

\vspace{3in}

$^*$Corresponding authors: sbelomes@fnal.gov, sposen@fnal.gov

\newpage

\tableofcontents

\newpage

\def\thefootnote{\fnsymbol{footnote}}
\setcounter{footnote}{0}

\section{Executive summary}

Over the last decades, the superconducting radio frequency (SRF) technology has made tremendous progress and is nowadays a critical technology for many high energy physics (HEP) accelerators and experiments \cite{Padamsee_book_vII, Belomestnykh2012,Padamsee50years}. In the near future, SRF linac PIP-II will provide H$^-$ beam to drive LBNF/DUNE experiment at Fermilab, SRF crab cavities will chirp the proton bunches at the ATLAS and CMS detectors of HL-LHC to boost the collider's luminosity. Further down the road, the SRF technology will be crucial for the success of a future Higgs factory, being either SRF linear collider option, such as ILC and HELEN, or circular collider option, such as FCC-ee and CEPC. Recently demonstrated excellent performance of SRF cavities in quantum regime (single-photon operation at millikelvin temperatures) enables new set of experiments in dark matter searches and new avenues for building quantum computers, which would be very beneficial for the HEP field. The new applications of the SRF technologies pose new challenges that must be addressed via a dedicated SRF R\&D program. There is still a vast room for improvements and many new ideas to explore.

A roadmap of SRF R\&D for the decade starting in 2018~\cite{GARD-RF-Strategy} was developed under the framework of the DOE GARD (General Accelerator R\&D) program. The roadmap was developed by a team of leading researchers in the field from national laboratories and universities, both domestic and international. It provides a community-directed guidance and reflects the most promising research directions for advances that enable future experimental high energy physics programs. In this contributed paper, we lay out a framework for key SRF R\&D directions, which are largely aligned with and extend beyond the GARD roadmap.

Among the key directions that we consider in this paper are:
\begin{itemize}
    \item studies to push performance niobium and improve our understanding of RF losses and ultimate quench fields via experimental and theoretical investigations;
    \item developing methods for nano-engineering the niobium surface layer and tailoring it for specific applications;
    \item investigations of new SRF materials beyond niobium via advanced deposition techniques and bringing these materials to practical applications;
    \item developing advanced cavity geometries to push accelerating gradients of bulk niobium cavities to $\sim70$~MV/m and pursuing R\&D on companion RF technologies to mitigate field emission, provide precise resonance control, etc.;
    \item research on application of SRF technology to dark sector searches.
\end{itemize} 

Strategic investments in these SRF R\&D areas would open opportunities to build new, more efficient, compact, and cost-effective HEP accelerators and enable new types of experiments. To realize these opportunities, we ask Snowmass 2021 for a strong recommendation to increase investment in the SRF research and technology development.

\section{Introduction: SRF cavity performance frontier}

Superconducting radio frequency (SRF) is a critical technology for several frontiers of experimental high energy physics. For example: SRF cavities make up the vast majority of the PIP-II linac, which will drive LBNF/DUNE \cite{LBNF-DUNE}; SRF cavities accelerate beams of the LHC, and they will also provide crabbing at the interaction regions to boost luminosity of the HL-LHC \cite{Apollinari:2017lan}; SRF cavities would provide energy for beams in the next generation of proposed Higgs factories, including ILC~\cite{ILC_TDR-v3-I,ILC_TDR-v3-II,ILC_Snowmass2021}, FCC-ee~\cite{FCC:2018byv}, and CEPC~\cite{CEPC_CDR}. In addition, SRF cavities are being explored not only for particle beam acceleration but for detection in the next generation of dark sector searches~\cite{braine2020,Tobar_LOI,DarkSRF,HarnikSearchesSRF-Snowmass2021} as well as for quantum computing, which could be extremely beneficial for HEP applications~\cite{Harnik_LOI,Matchev_LOI}. To continue enabling future high energy physics experiments, research and development on SRF cavities is crucial. Continued improvements in cavity performance make new scientific applications feasible whereas they would have otherwise been either un-achievable or too expensive. 

\begin{figure}
\begin{center}
\includegraphics[width=0.75\textwidth]{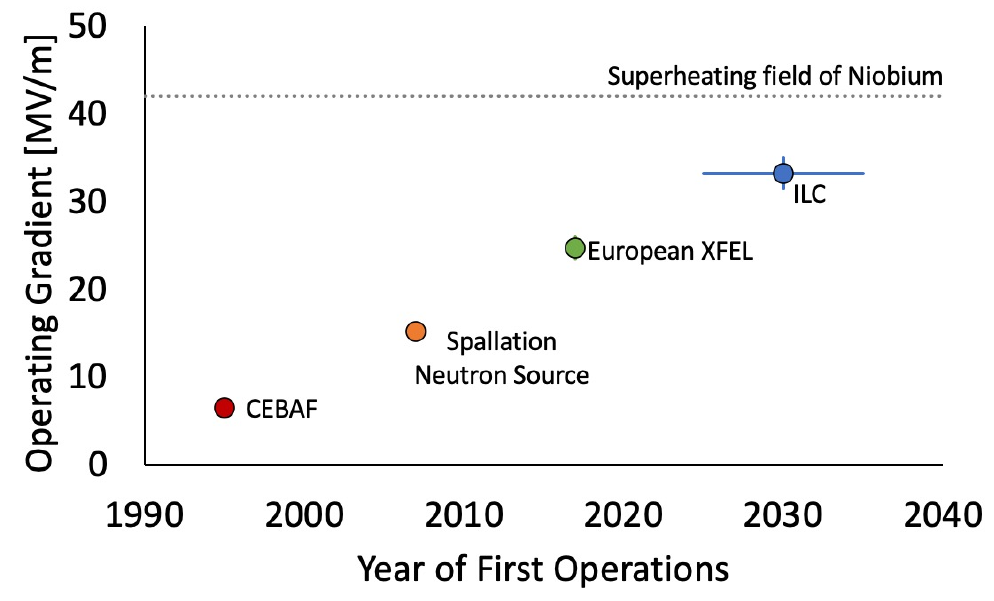}
\end{center}
\caption{Progress with time in SRF R\&D leading to higher operating gradients for large SRF accelerators.}
\label{fig:SRF_progress}
\end{figure}

Fig.~\ref{fig:SRF_progress} illustrates how SRF R\&D can enable experimental physics. It plots operating gradient of large-scale SRF linear accelerators (order of hundreds of SRF cavities) versus year of first operations. It shows how as progress was made in SRF R\&D to increase accelerating gradient -- through years of efforts to ameliorate degradation mechanisms such as multipacting, thermal breakdown, and field emission~\cite{KleinProch1979,koechlin1996,KneiselSRF1993,Padamsee_History} -- the new capabilities would bring into reach new accelerator-based experimental programs in nuclear physics, basic energy sciences, and high energy physics.

Fig.~\ref{fig:SRF_progress} includes ILC, based on gradient targets from the TDR~\cite{ILC_TDR-v3-I,ILC_TDR-v3-II}. In a Snowmass 2021 submission~\cite{ILC_Snowmass2021}, there is a discussion of how the baseline gradient could be significantly increased in several upgrade scenarios based on continous R\&D progress. The figure does not show other ways in which R\&D has improved cavity performance, such as improvements in $Q_0$ (e.g., the factor of $\sim3$ improvement in $Q_0$ provided by nitrogen doping has been an enabling factor for the LCLS-II X-ray FEL project~\cite{N-doping,LCLS-II_FDR}; and $Q_0$ values $>3\times10^{11}$ at temperatures ~1.4~K have been achieved by medium temperature baking~\cite{Posen_mid-T_baking}). These and  new advances in the near future would be crucial for new proposals, e.g., HELEN collider~\cite{HELEN}.

\begin{figure}
\begin{center}
\includegraphics[width=0.85\textwidth]{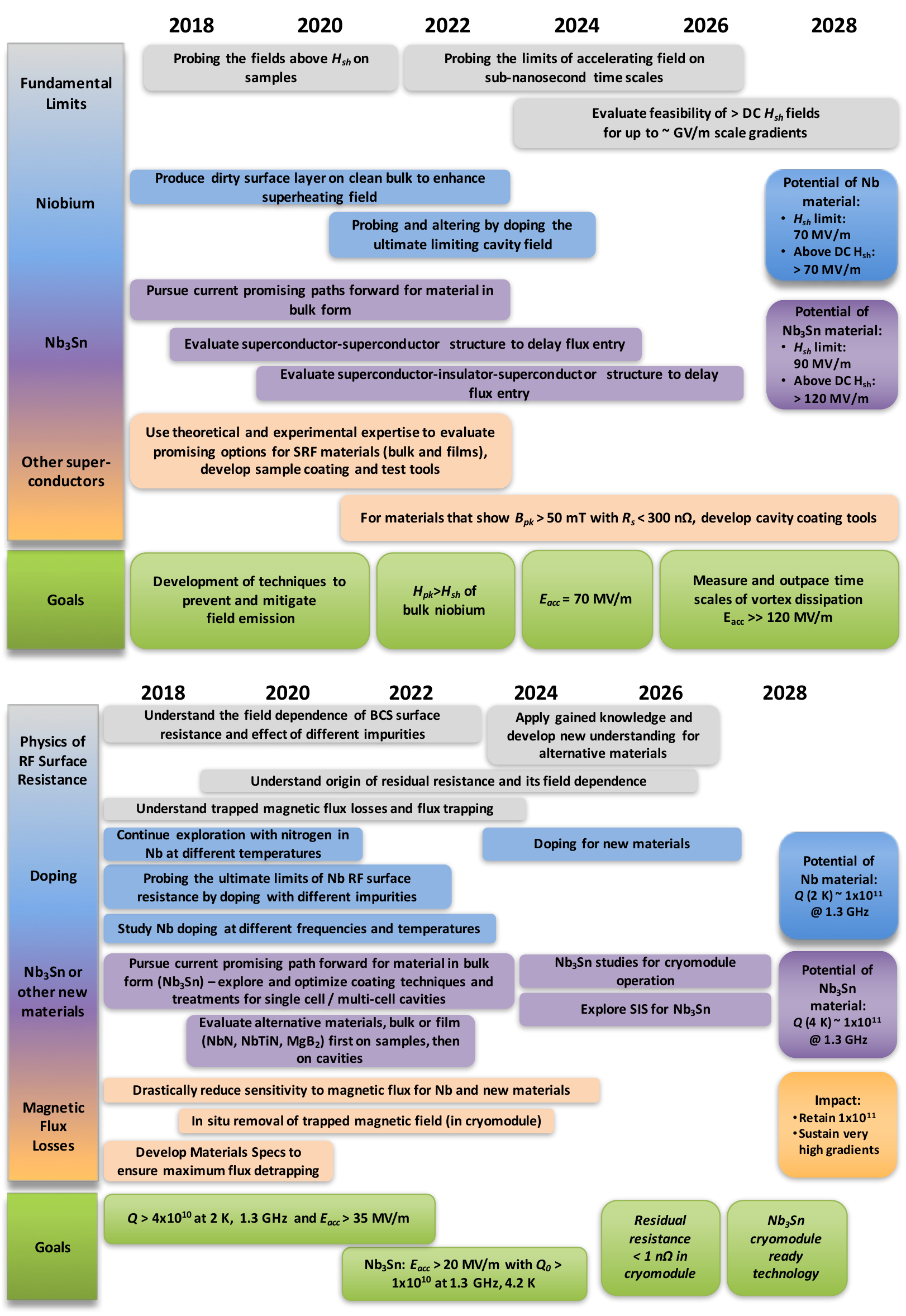}
\end{center}
\caption{Tables showing proposed directions over a decade of research and development towards pushing the accelerating gradients (top) and $Q_0$ (bottom) of SRF cavities (from~\cite{GARD-RF-Strategy}).}
\label{fig:GARD_roadmap}
\end{figure}

A roadmap of SRF R\&D for the decade starting in 2018 was developed under the framework of the DOE General Accelerator R\&D (GARD) program. The roadmap was developed by a team of leading researchers in the field from national laboratories and universities, both domestic and international. The GARD roadmap for SRF R\&D provides community-directed guidance and reflects the most promising research directions for advances that enable future experimental high energy physics programs. Fig.~\ref{fig:GARD_roadmap} shows two summary tables from the 2017 road-mapping exercise~\cite{GARD-RF-Strategy}. Recently, the 2020 update of the European Strategy for Particle Physics~\cite{ESPP2020} emphasized the importance of accelerator R\&D. The followed up effort on developing a roadmap for European accelerator R\&D culminated in a report, which includes a roadmap for high-gradient RF structures and systems~\cite{ESPP-RF-roadmap}. The European roadmap covers SRF cavity topics similar to the GARD roadmap, and hence there is a lot of synergy and opportunities for collaboration between the U.S. and European laboratories and universities. 

In this contributed paper, we outline our views on key research directions, which largely continue to align with the GARD roadmap. These include studies pushing the performance of niobium, including doping, multi-step heat treatment, flux expulsion and flux losses. The key directions include new materials such as Nb$_3$Sn as well as layered structures. They also include fundamental studies of the physics of RF surface resistance and penetration of flux into superconductors at high fields. Further, the paper discusses advanced SRF cavity shapes, use of SRF cavities for dark sector searches, and other key R\&D directions. The goal is to highlight the importance of the SRF cavity technology for future HEP accelerators and experiments and propose an update to the GARD roadmap.

\section{Nano-engineering the Nb surface layer for high Q and high gradient}

Superconducting properties of niobium cavities -- surface resistance and breakdown magnetic field -- depend on the state of material in the first few tens of nanometers of the RF surface of the cavities. For example, the penetration depth of a clean niobium is about 40~nm. Only relatively recently the techniques have been developed to systematically change the surface properties and study the effects of the changes, e.g., using HF rinse~\cite{Romanenko2013}. The ultimate goal of the experimental studies -- in parallel with developing theoretical concepts and models~\cite{GurevichSnowmass2021} -- is to come up with methods for nano-engineering the surface layer to control its properties for optimal, project-dependent SRF performance. As the cavity performance is one of the main cost driver for SRF accelerators, such “tailored surface” methods would offer prospects for a dramatic reduction of accelerator footprint, construction and operation costs, thus broadening the range of applications.

Improving $Q$ factors of niobium cavities at medium gradients (16--25~MV/m) is a key R\&D area for future CW and high duty factor HEP accelerators, such as circular $e^+e^-$ colliders FCC-ee~\cite{FCC:2018byv} and CEPC~\cite{CEPC_CDR}, a CW SRF linac for PIP-II upgrade~\cite{8GeVSRFlinac}, etc. This key area is synergistic with rapid progress over the past 5--10 year in developing nitrogen doping technology~\cite{N-doping} for SRF-based free electron lasers, specifically LCLS-II and LCLS-II-HE. The most recent nitrogen doped 1.3~GHz cavities for LCLS-II-HE demonstrate an average $Q$ of $3.6\times 10^{10}$ at 23~MV/m~\cite{Gonnella_TTC2022}. 
Presently, it looks like the most promising path to even higher quality factors would come from mid-temperature ($\sim300^{\circ}$C) baking of SRF cavities in vacuum, which dissolves the natural oxide and other surface layers into the bulk of niobium~\cite{Posen_mid-T_baking}, with demonstrated $Q = 5\times10^{10}$ at 30~MV/m. While the $Q$ factor drops after exposure to air and High-Pressure Rinsing with ultra-pure deionized water (HPR), the cavities still demonstrate quite good performance with $Q$ in the range of $3.5 - 4.4\times10^{10}$ at gradients between 16–-24~MV/m~\cite{he2020mediumtemperature}. This is a very promising and active R\&D area. The name of the game in improving $Q$ is improving the cavity surface resistance. While some progress was made, the main open questions remain as it was highlighted in~\cite{GARD-RF-Strategy}. To answer these questions, the SRF community should:

\begin{itemize}
    \item continue exploration of the effect of interstitial impurities on bulk Nb surface resistance;
    \item develop fundamental understanding of the reverse field dependence of the BCS surface resistance and devise experiments towards validation of different theories;
    \item develop understanding of mechanisms of trapping magnetic vortices and their contribution to the RF losses, and devise experiments towards validation of models;
    \item develop understanding of `intrinsic' residual resistance and its field dependence;
    \item ameliorate trapped vortices via innovative ideas: advanced magnetic shielding concepts, in situ flux removal, developing material properties/preparation for minimal pinning strength, etc.
\end{itemize}

Pushing toward higher gradients while improving cavity quality factors is closely tied with investigating fundamental electromagnetic field limit in SRF cavities. There was some recent empirical progress. Two low-temperature surface treatment recipes demonstrated accelerating gradients higher than the standard ILC treatment. First, the nitrogen infusion~\cite{nitrogen_infusion} process demonstrated gradients of 40--45~MV/m in 1.3 GHz TESLA-shaped cavities. Second, near 50~MV/m have been achieved with a $75/120^{\circ}$C two-step bake with cold electropolishing treatment~\cite{2-step_baking,Bafia_SRF2019}, which is a more robust recipe than infusion. However, there is still no good theoretical model explaining what the dynamic superheating field $H_{sh}(\omega,T)$ is~\cite{GurevichSnowmass2021}. It is very important to continue theoretical and experimental research to establish the ultimate SRF field of niobium.

\section{Nb$_3$Sn as practical Nb material}
\label{sec:Nb3Sn}

Nb$_3$Sn superconductor has potential of reaching fields approximately twice as high as niobium, or $E_{acc}\sim$~100~MV/m ~\cite{Catelani2008,Lin2012,Kubo2020}. Extrapolation from high power pulsed RF experiments is in agreement with this prediction~\cite{PosenPRL}. In addition, having two time higher critical temperature compared to niobium, Nb$_3$Sn cavities should have quality factors at $\sim$~4~K similar to niobium cavities at 2~K. This would lead to a significant cryogenic power advantage in large scale applications, e.g., see \cite{Posen_PoP-demonstration}. Nb$_3$Sn cavity performance has not reached its ultimate performance potential yet, but it has been making substantial progress. In CW regime, Nb$_3$Sn deposited on bulk niobium cavities show promising results~\cite{PosenSnowmass2021}, reaching high $Q\sim10^{10}$ at 4.4~K. So far, R\&D cavities demonstrated accelerating gradients up to 24~MV/m in single-cell cavities and 15 MV/m in 9-cell TESLA-shape cavities \cite{Posen_2021}. Such gradients might already be useful for small-scale accelerators. In addition, Nb$_3$Sn SRF cavities could have an advantage for dark matter searches due to the material ability to remain superconducting in large magnetic fields thus outperforming copper cavities that are used in such experiments so far. For example, initial results are quite promising, demonstrating $Q_0$
of $\sim5\times10^5$ in DC magnetic field of 6~T at 4.2~K for a special shape cavity~\cite{PosenNb3Sn-magn-field}.

For accelerator-based HEP applications, further developments are necessary. Circular $e^+e^-$ colliders FCC-ee~\cite{FCC:2018byv}, CEPC~\cite{CEPC_CDR} and other CW/high duty factor accelerators (e.g., a CW SRF linac for PIP-II upgrade~\cite{8GeVSRFlinac}) would require medium gradient (in the range 15--25~MV/m), high efficiency multi-cell cavities. Pulsed linacs for an energy upgrade of ILC~\cite{ILC_Snowmass2021} or for HELEN collider~\cite{HELEN} would need very high gradient cavities of $\geq80$~MV/m that can still maintain high efficiency.

Due to relatively short coherence length $\sim3$--4 nm, approximately an order of magnitude smaller than niobium, the Nb$_3$Sn cavity performance is more sensitive to surface defect. There are experimental results indicating that indeed this is the case~\cite{PosenPRL,PorterThesis}. As the post-coating surface treatments are not well developed yet, more R\&D efforts in this area are in order. In parallel to bulk Nb$_3$Sn film R\&D, alternative manufacturing routes should be pursued as well. For example, thin Nb$_3$Sn films on clean Nb or Superconductor-Insulator-Superconductor (S-I-S)~\cite{Gurevich2006,Kubo2021} Nb$_3$Sn layered structures should be studied as potential ways to achieve the superheating field of Nb$_3$Sn. These efforts would require more development to ensure good RF performance. A white paper~\cite{PosenSnowmass2021} outlines the current status of the Nb$_3$Sn research and proposed R\&D directions in more details.

\section{Other materials beyond bulk Nb}

Superconductors other than bulk niobium have been used in non-RF applications for many years, see for example~\cite{Nb3SnMagnets}. However, it is still very difficult to apply those to SRF cavities due to extremely high sensitivity of surface resistance to the quality of deposited films. At present, the only material close to become practical for SRF cavities is Nb$_3$Sn, discussed in the previous section. 

Technology of depositing thin films of niobium on copper (Nb/Cu) has been used in accelerators that do not require high gradients and can operate at 4.5~K (that is at low RF frequencies), such as LEP2, LHC, and HIE-ISOLDE at CERN, ALPI at INFN-LNL (Italy), SOLEIL (France). Although exhibiting very high $Q$ at low field, sputtered Nb on Cu cavities have been plagued with steep increase of RF losses at medium to high gradients. Recent R\&D efforts to develop advanced coating methods have produced some encouraging results (see references in~\cite{ValenteSnowmass2021}), but further efforts are needed to scale up the processes and firmly establish the good baseline performance of SRF cavities.

Alternative superconducting materials with higher critical field and critical temperature can potentially surpass the intrinsic limits of niobium. Among the advantages the new materials might offer are i) high $Q$ at temperatures above 2~K for higher efficiency and lower cost of large-scale facilities and small scale cryocooler-based accelerators, and ii) higher accelerating gradients for reaching higher beam energies in more compact accelerators.

Among the alternative materials that are still being evaluated are A15 superconductors (in addition to Nb$_3$Sn,  those are Nb$_3$Al, V$_3$Si, ...), MgB$_2$, pnictides, etc.~\cite{Valente2016}. All this materials require development of deposition techniques on a suitable for SRF cavities substrate, e.g., copper or niobium. Deposition methods such as sputtering, energetic condensation and atomic layer deposition (ALD) are being explored. Thin films could also bring the advantage of layered structures (e.g., S-I-S) for a potential to further boost achievable accelerating gradients. R\&D to improve deposition methods must continue so that the most promising techniques are selected, and alternative superconductors with practical levels of surface resistance are applied to SRF cavities for performance evaluation in a reasonable time frame.

\section{Advanced SRF cavity shapes}

For an $e^+e^-$ linear collider, the 1.3~GHz TESLA cavity~\cite{TESLAcavity}, along with other aspects of SRF technology, was developed in 1990's. While the cavity has an optimized geometry and served the community well enabling several small and large accelerators (the prime examples are European XFEL~\cite{E-XFEL} and LCLS-II/LCLS-II-HE)~\cite{LCLS-II,LCLS-II-HE}), there is room for improvement of the geometry to reduce the ratio of surface magnetic field $H_{pk}$, which is fundamentally limited by the field at which superconductivity breaks down, to accelerating gradient $E_{acc}$. A reduction of $H_{pk}/E_{acc}$ by 10--20\% can be achieved. Several advanced cavity geometries have been proposed over the years: Re-entrant~\cite{Re-entrant_cavity, Optimal_cells}, Low-Loss \cite{Low-Loss_cavity}, ICHIRO~\cite{ICHIRO_cavity}, and Low Surface Field (LSF)~\cite{Li_LSFcavity}. Many single cell R\&D cavities of these geometries were built and tested with the standard ILC surface treatment, demonstrating gradients well above 50~MV/m. Combining one of these cavity geometries with the most advanced surface treatments, such as two-step low-temperature bake, one can expect improving gradients up to 60~MV/m in 9-cell standing wave structures. However, very limited efforts have been devoted to multi-cell cavities so far. More funding for R\&D in this area could provide rapid progress in a short time. More details on the advanced cavity shapes can be found in~\cite{HELEN} and references therein. An advanced cavity shape could be used for an energy upgrade of ILC~\cite{ILC_Snowmass2021} or for a more compact Higgs factory HELEN~\cite{HELEN}.

Travelling wave (TW) SRF structures offer further advantages over standing wave: substantially lower peak magnetic ($H_{pk}/E_{acc}$) and lower peak electric field ($E_{pk}/E_{acc}$) ratios, two times higher $R/Q$. The TW shape optimization~\cite{TW_optimization} indicates that an accelerating gradient as high as 70~MV/m might be achievable. The 38\% increase of the $G\cdot R/Q$ parameter reduces the cryogenic dynamic heat load at high gradients. The higher  cell-to-cell coupling of the TW mode makes the structure less sensitive to cavity detuning errors, making tuning easier, despite the larger number of cells. Also, high stability of the field distribution along the structure with respect to geometrical perturbations allows for longer accelerating structures (compared to TESLA cavities), limited by manufacturing technology.

Developing novel fabrication methods could enable the use of structures considered previously only for NCRF systems, for example a parallel-feed accelerating structure~\cite{Welander2017}. If a robust, high-performance coating of thin-film niobium (or other superconductor) on copper is developed, such structures could potentially be more efficient for future accelerators, both linear and circular.

Future HEP accelerator-based experiments require higher beam intensities and/or shorter synchrotron cycle times. The higher beam intensities of future circular colliders mean that the accelerating cavities should incorporate strong damping of higher order modes (HOM). While such cavities have been successfully developed and operated, see e.g.,~\cite{Belomestnykh2012}, new approaches are always sought to improve the performance even further. Some machines require low-frequency RF, for which quarter-wave, $\beta=1$ resonators have been developed~\cite{56MHz,28MHz,Zhang2021}. However, proton synchrotrons, either new or upgraded existing (e.g., the Main Injector at Fermilab) require fast frequency tuning during the machine cycle. The SRF technology would significantly reduce the number of cavities in these machines, thus reducing parasitic impedance due to accelerating structures, but the lack of fast frequency tuners prohibits the use of SRF cavities at present. Development of the required fast frequency tuners (we discuss the fast tuner R\&D further in section~\ref{sec:Other_directions}) in parallel with new cavity structures operating at $\sim50$~MHz would provide much higher acceleration gradients and; therefore, a smaller number of cavities and lower the beamline impedance.

\section{SRF for dark sector searches}

SRF cavities are resonators with extremely high quality factors. As such, they are of a strong interest for quantum information science and experiments searching for dark-matter particles. In particular, SRF cavities can serve as an extremely sensitive detectors of very weak signals when new particles (e.g., dark photons or axions) or high-frequency gravitational waves convert to microwave photons~\cite{HarnikSearchesSRF-Snowmass2021}. 3D SRF cavities provide large volumes where the photons can be collected with high efficiency, that is the photons would have a very long lifetime due to extremely low losses in the cavity walls.

While some salient features that make SRF cavities so attractive for particle accelerators (high acceleration rate, high $Q$ at medium and high gradients) are not relevant for most experiments discussed here, maintaining high quality factor at very low electromagnetic fields and millikelvin temperatures is of utmost importance. Experiments with 2D and 3D resonators show that the $Q$ factor goes down at very low temperatures and electromagnetic fields. This is an indication that the two-level systems (TLS) residing inside an amorphous niobium oxide layer may play a significant role in the low-field performance~\cite{TLS-Anderson,TLS-Martinis,TLS-Romanenko}. Reducing TLS losses is an active area of research~\cite{Romanenko:2018nut}, but more efforts are required to find a robust method to consistently achieve high $Q$ in quantum regime. In addition, there is a need to achieve high $Q$ when a cavity is in a multi-tesla magnetic field.

Traditional, TESLA-shape SRF cavities can be used as is in experiments such as Dark SRF (light shining through wall)~\cite{DarkSRF2}. However, many other experiments would require special cavity geometries and/or operating in high DC background magnetic field, which we mentioned in section~\ref{sec:Nb3Sn}. For example, a cigar-shaped cavity~\cite{PosenNb3Sn-magn-field} for axion searches was developed specifically to reduce losses in Nb$_3$Sn due to external DC magnetic filed. Another proposed method for axion search is to employ two modes of a cavity with such a field configuration so that an RF magnetic field of the second mode is used in lieu of the DC magnetic field, see details in~\cite{HarnikSearchesSRF-Snowmass2021} and references therein. A special two-mode cavity will have to be developed for this experiment.

\section{Other key R\&D directions}
\label{sec:Other_directions}

There are other research areas and companion technologies that are important for success of SRF cavities in different operating regimes. Here we mention just a few of them.

\noindent
\textbf{Mitigation of field emission}: Imperfections of the cavity inner surface or contamination by dust particles enhance local electric field and can cause emission of electrons from the cavity surface at high gradients. This so-called field emission (FE) could be a serious impediment to achieving high gradients. To reduce the probability of emitters to appear, special care should be taken during the cavity preparation and assembly. In particular, all SRF cavities are high-pressure rinsed with ultra-pure deionized water and assembled in clean rooms of Class 100 or better. Even when using the state-of-the-art techniques, some cavities still suffer from FE. Special studies must be performed in parallel with high gradient research to abate FE. A promising pathway is to develop robotic assembly of SRF cavities and cryomodules to eliminate human errors and reduce sources of possible contamination~\cite{BerryTTC2019,GuoTTC2020}. This is still a nascent area. It requires better funding to produce results in reasonable time. Other post-assembly in situ methods that require further development include, e.g., plasma processing~\cite{MartinelloSnowmass2021}.

\noindent
\textbf{Compensation of microphonic noise and Lorentz force detuning}: Advances in the SRF cavity performance should be accompanied by R\&D on the cavity resonance control~\cite{SchappertLINAC2016,PischalnikovIPAC2015}.
SRF cavities operating in CW regime, especially with light beam loading like energy recovery linacs and low-beam-current proton linacs (and hence with high loaded $Q$'s) are susceptible to large swings of the resonant frequency due to vibrations of their walls. These vibrations are caused by external excitation from different sources and are commonly known as microphonic noise, or simply microphonics. Compensation of microphonics becomes particularly important for operations at $\sim4$~K (e.g., for Nb$_3$Sn cavities): liquid helium systems are inherently noisier at temperatures above the superfluid transition temperature 2.17~K. For pulsed linacs, such as European XFEL, ILC, and HELEN, compensating cavity resonant frequency detuning due to the Lorentz force (Lorentz Force Detuning or LFD) is especially important as gradients of the linac increase, because the ratio of LFD to the cavity bandwidth is proportional to the cube of accelerating gradient. R\&D efforts should focus on i) new cavity designs optimized for response to LFD and microphonics, and ii) new active LFD and microphonics compensation algorithms along with the new tuner (fine and coarse) designs. 

\noindent
\textbf{Ferroelectric tuners}: Ferroelectric ceramic materials with low loss tangent at RF frequencies~\cite{Nenasheva2010,Kozyrev2009,KanareykinIPAC2010} allow development of electrically controlled devices (tuners) with much shorter switching times than those of piezo-electric mechanical tuners. Such a ferroelectric tuner, when inserted into a high-power transmission line connected to the SRF cavity, can allow to do two things: i) alter the coupling between the transmission line and the acceleration structure~\cite{Kazakov2006}, and ii) electronically control the cavity frequency within a bandwidth needed for active compensation of microphonics. A ferroelectric tuner could eliminate the need for overcoupled fundamental power couplers, thus significantly reducing RF amplifier power. A proof-of-principle demonstration of the ferroelectric fast reactive tuner (FE-FRT) was recently accomplished at CERN~\cite{ShipmanSRF2019}. Further R\&D efforts are needed in this area to realize the full potential of ferroelectric devices.

\section{Summary and conclusions}

Superconducting radio frequency cavities have improved dramatically in performance over the last decades, thanks to dedicated R\&D efforts. These improvement have enabled new scientific applications that had not previously been feasible. Even considering these significant past improvements, there is still vast room for continued improvement through additional R\&D. These efforts are funding-limited, not idea-limited, and more ideas continue to be generated, as more experimental advances are made. The exciting areas to explore, that were discussed in this document, include 
\begin{itemize}
    \item furthering our understanding of RF losses and ultimate quench fields of niobium via experimental and theoretical investigations;
    \item developing methods for nano-engineering the niobium surface layer and tailoring it for specific applications;
    \item studying new SRF materials beyond niobium via advanced deposition techniques and bringing these materials to practical applications;
    \item developing advanced cavity geometries to push accelerating gradients of bulk niobium cavities to $\sim70$~MV/m and pursuing R\&D on companion RF technologies to mitigate field emission, provide precise resonance control, etc.;
    \item investigating application of SRF technology to dark sector searches.
\end{itemize} 

Strategic investments in these R\&D areas would open opportunities to build new, more efficient, compact, and cost-effective HEP accelerators and enable new types of experiments. To make these performance improvements realizable, we ask Snowmass 2021 for a strong recommendation to increase investment in the SRF research and technology development.

\section{Acknowledgements}

Work supported by the Fermi National Accelerator Laboratory, managed and operated by Fermi Research Alliance, LLC under Contract No. DE-AC02-07CH11359 with the U.S. Department of Energy. The U.S. Government retains and the publisher, by accepting the article for publication, acknowledges that the U.S. Government retains a non-exclusive, paid-up, irrevocable, world-wide license to publish or reproduce the published form of this manuscript, or allow others to do so, for U.S. Government purposes.





\bibliographystyle{JHEP}

\bibliography{myreferences} 






\end{document}